# High-precision radial-velocity measurement with a small telescope: Detection of the tau Boötis exoplanet

Thomas G. Kaye[1], Sigfried Vanaverbeke[2,3], John Innis[4]

[1] Spectrashift.com, 404 Hillcrest, Prospect Heights IL 60090, USA
[2] KULAK, E.-Sabbelaan 53, 8500 Kortrijk, Belgium
[3] Vereniging voor Sterrenkunde (VVS), Brieversweg 147, 8310 Brugge, Belgium
[4] 280 Brightwater Rd, Howden, Tasmania, 7054, Australia

The successful detection is reported of radial-velocity variations due to orbital motion of the substellar companion of the star tau Boötis, from data obtained with a small aperture (0.4m) telescope and a fibre-fed high-resolution spectrograph. Radial-velocity observations from observing runs in 2000 and 2004 reveal a periodic variation of $3.30 \pm 0.02$d, which is consistent with the previously determined value of $3.3125 \pm 0.0002$d. We fit our data to a circular orbit with the known period, and derive a velocity amplitude of $471 \pm 10$m s-1 (in agreement with the previously published value of $469 \pm 5$m s$^{-1}$), and determine a time of maximum velocity (Tmax) of HJD $2453113.95 \pm 0.01$. These observations explore the minimum system requirements for precise radial-velocity measurements.

## Introduction

The state of the art in very precise radial-velocity measurements of solar-type stars has reduced the noise level to the 1m s$^{-1}$ range, e.g. Pepe *et al.*(2002).[1] This is accomplished with high-quality spectrographs on some of the world's best and largest professional telescopes. However, as many of the extrasolar planets have orbital velocity amplitudes of a few times 100m s$^{-1}$ or more, these should be detectable with more modest, albeit specialized equipment. We report the detection of the orbital motion of tau Boötis, arising from the presence of an extrasolar planet, using radial-velocity data obtained with amateur equipment.

Precise radial-velocity (Doppler) monitoring of solar-type stars has been the most successful planet-detection method used to date, and accounts for more than 90% of the discoveries.[2,3] A total of 161 planet candidates were known at the start of 2005 July, and their numbers are steadily growing. For the most recent information please see the excellent Extrasolar Planets Encyclopedia maintained by Jean Schneider at http://www.obspm.fr/-encycl/encycl.html. These newly discovered systems support some of the features predicted by theories of star and planet formation, but systems with massive planets having very small orbital radii and eccentric orbits appear common, and were generally unexpected.[2,4]

Recent developments in commercial CCDs, software tools and telescopes enable individual amateur astronomers and small research institutions to develop capabilities which, even only a few years ago, were thought to be entirely beyond their reach. Organizations such as the American Association of Variable Star Observers (www.aavso.org), the Center for Backyard Astrophysics (CBA, http://cba.phys.columbia.edu) and Tim Puckett's supernova search team (http://www.cometwatch.com) are demonstrating such capability on a regular basis. However, spectroscopy is an area where small-telescope users have lagged behind. The velocity amplitude which is induced by an extrasolar planet on its parent star is generally much less than 1 km s$^{-1}$.[5] Spectrographs capable of very precise measurements, with excellent long-term stability, are required to detect such small velocity variations. The technical requirements for the telescope are less stringent, and mainly require enough aperture to be able to record spectra with sufficient signal-to-noise ratio, although a highly accurate and stable guiding system is needed. With this as motivation, one of us (TK) began construction of a series of medium to high-resolution spectrographs using readily available components, and commenced a study of the factors that would limit velocity precision when using a small commercial telescope.

The star tau Boötis (V=4.5) is one of the brightest stars in the sample of known planet-bearing main-sequence stars. The substellar companion of this star was discovered by Butler *et al.*[6] in 1997, although Duquennoy & Mayor (1991)[7] had earlier reported evidence for radial-velocity variations. The brightness of the parent star and the relatively large radial-velocity amplitude of the extrasolar planet indicated that a small (~0.5m) telescope should be able to detect the planet, given a sufficiently stable spectrograph and a sufficiently long series of observations. Hence this star was chosen as a 'proof of concept' test for the spectrograph. This paper presents radial-velocity data obtained in 2000 and 2004 with a 0.4m telescope. Spectral and least-squares analysis of the data show clearly a periodic variation consistent with the known properties of this extra-solar planetary system. The use of amateur equipment to detect the planet orbiting tau Boötis demonstrates significant progress in this field, and opens up new possibilities for radial-velocity studies using equipment built from off-the-shelf components.

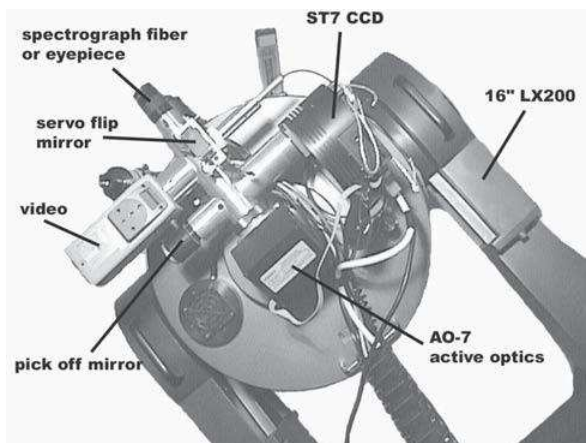

**Figure 1.** Rear view of the Meade 16" (400mm) LX200 telescope showing the components used for high-precision guiding.

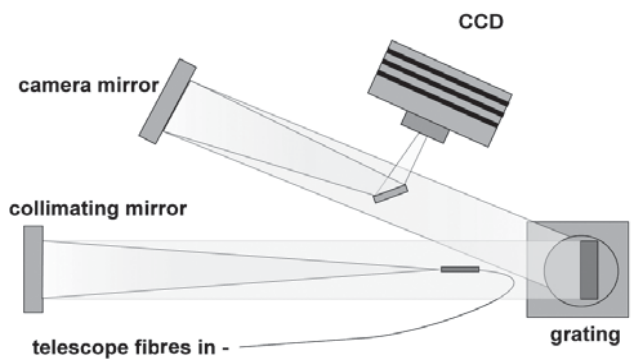

**Figure 2.** Basic layout of the Czerny–Turner spectrograph. Optical fibres are mounted on a thin vane to minimise obstruction. Mirrors are 15cm diameter parabolic in granite mounts. All components are mounted on a granite slab.

## Equipment

### Telescope and acquisition/guiding system

The telescope used is a Meade 16" (400mm) LX200 (Figure 1). The focusing system was re-machined to minimise primary-mirror movement, otherwise the unit was standard. An AO7 tip/tilt mirror from Santa Barbara Instruments Group (SBIG) was positioned on a custom machined mount. The mount incorporates a pick-off mirror to direct a small percentage of the target star's light to the tracking chip in an SBIG ST-7 CCD camera. The AO7 and ST-7 monitor and correct the position of the star 8 times a second through actuation of the AO7 mirror. The main beam of starlight is brought to a seven-strand fibre bundle at the Cassegrain focus. The fibre bundle allows for easier acquisition of the star and once the light is detected in one fibre a simple adjustment of the AO7 is made to centre the star on the central fibre. The star is held centrally on that fibre for the duration of the exposure by the use of the AO7/ST-7 combination.

The use of fibre optics is the best method to connect a small telescope to a large spectrograph because optical fibres can be run over distances of many tens of metres with little loss in throughput, allowing the spectrograph to be remotely located on a bench in a controlled environment. Although the fibre provides a measure of image scrambling, changes in the illumination of the fibre can result in small changes in the illumina-tion of the spectrograph optics, and hence small variations in the spectra. The AO7 helps reduce guiding errors induced by 'seeing noise' which would otherwise track through to increase the errors in the velocity measurement.

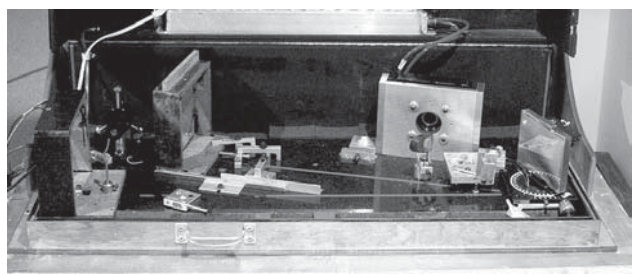

**Figure 3**. Photograph of the spectrograph. The chassis is a 60×90cm black granite slab drilled for threaded inserts at strategic locations. The mirrors are mounted on the left, the diffraction grating on the lower right, and the CCD is on the upper right.

### Spectrograph

The spectrograph is a typical Czerny–Turner design with a fibre-optic feed. The basic layout of the spectrograph and a photograph of the system are shown in Figures 2 and 3. The fibre bundle directs starlight to a 762mm focal-length collimating mirror. The collimated beam illuminates the plane grating (1800 lines per mm), and a 240mm focal-length camera mirror directs the beam to an Apogee AP7 CCD. The Apogee camera has a thinned back-illuminated CCD with very high efficiency, especially in the blue region of the spectrum.

The dispersion of the spectrum on the CCD is approximately 0.17Å/pixel near the operating wavelength of ~5100Å. The CCD is 512 pixels wide, giving a spectral coverage of about 88 angstroms. In our instrument the fibre end forms the spectrograph 'slit' which, when imaged on the CCD, has a full width at half maximum (FWHM) of 4 pixels or 0.68Å. Hence the resolution of the spectrograph, $\lambda/d\lambda$, where $\lambda$ is the wavelength and $d\lambda$ is the FWHM, is about 7500. Professional planet detection spectrographs generally use resolutions between 30,000 and 60,000. For comparison, the resolution of most commercial 'amateur' spectrographs is only a few hundred.

Maximum environmental and mechanical stability are the primary goals in the construction of a spectrograph that needs to measure spectrum shifts of the order of 1 micron on the detector. The spectrograph components are mounted on a 31mm-thick granite base, which is supported on a high-density foam pad. The mirror components are also mounted on specially made granite supports. The entire assembly is sealed in a double-insulated wood and foam box, in a temperature-controlled room. A small recirculating chiller supplies cooling water to the CCD and radiator inside the box. The chiller system is stabilised at about 16°C. A small electric heater is thermostatically controlled to raise the temperature in the spectrograph to about 18°C. Temperature stability is achieved after about one day, and typically holds within one degree. During the observing run the temperature of the spectrograph was continuously recorded and never varied by more than 2°.

A reference spectrum is required to compensate for residual thermal and mechanical drift in the system. A thorium-argon hollow-cathode lamp is used. The reference lamp is focused on two optical fibres that are arranged with the stellar fibres going into the spectrograph, and yield comparison-arc spectra that lie in rows above and below the stellar spectrum on the CCD, and are obtained simultaneously with the stellar data.

### Software

The telescope, AO7 and spectrograph CCD are all run simultaneously with *Maxim DL*. This versatile commercial software package allows for control of the CCD camera while simultaneously correcting telescope-drive errors and coordinating the tip/tilt mirror of the AO7. *Maxim DL* has the useful ability to lock the tracking of a star to a particular pixel on the ST-7 tracking CCD. The target pixel on the tracking CCD is adjusted to correspond with the image of the star centred on the input fibre at prime focus.

The acquired spectra are post-processed with the *IRAF* software package, developed and distributed by the US National Optical Astronomical Observatories, to determine the velocity shift. *IRAF* offers a vast array of software tools to handle almost every aspect of astronomical data and image processing. The raw spectra are stored as FITS files, with two reference (thorium-argon) spectra and one stellar spectrum in each file. The *IRAF* package *DO3FIBRES* is set up using pixel apertures to extract each spectrum, which is typically 4 pixels tall and 500 pixels wide. *DO3FIBRES* formats each spectrum into a single line of data by binning each 4-pixel stack. The three single-line spectra are combined into a specialised *IRAF* multi-spectrum file. Once the multi-spectrum files are completed, the package *FXCOR* is used to measure the shifts by Fourier cross correlation. The software computes the Fast Fourier Transform of each spectrum and compares it to a chosen template spectrum, correcting for the Earth's rotation and orbital motion. The interested reader can find a more detailed description of our hardware and reduction pipeline in Kaye (2003).[8]

## Observations

### General

A total of 713 observations of tau Boötis was obtained from Winer Observatory in Arizona during three separate runs in 2000 and 2004. Our initial observing run in 2000 February and March was plagued with technical and weather problems but nevertheless provided 127 measured radial velocities over a run of 18 days' duration. Owing to the adverse weather conditions during that run some observations were of marginal quality. However, analysis of the data strongly indicated the detection of the radial-velocity variation arising from the presence of the known extrasolar planet of tau Boo. Spectral and other analysis showed definite evidence for a periodic variation at ~3.3d, and around 450m s$^{-1}$ in amplitude.

From these preliminary results it was determined that additional observations would be required to constrain the orbit of the planet accurately. Consequently two more runs were scheduled in 2004 April and May, of 12 and 11 days' duration respectively. These observing runs were scheduled during the spring season in Arizona because of the generally optimal weather conditions and the prolonged visibility of tau Boötis at that time of the year. The 2004 April and May observing runs provided 586 additional data points, which resulted in a much better coverage of the orbit. Replacement of a defective grating in the instrument after the 2000 run also meant that the 2004 data were of higher quality.

### Observing procedures

Each night's run is started by equalising the telescope temperature, adjusting focus and acquiring the target star. The Th/Ar lamp is turned on at least an hour before the run starts for stabilisation. The spectrograph is equipped with a photomultiplier tube (PMT) to monitor the stellar flux entering the fibres. Several pixel tracking coordinates are tested for best throughput and finally the PMT counts are used to determine exposure times (typically 15–20 minutes) to reach 100/1 SN for the spectrum. Post-processing subtraction of dark frames was handled by the *MaximDL* software using master darks compiled from fifty frames. The star is tracked all night and exposure times are adjusted periodically for seeing and air mass. Spectrograph temperature, ambient temperature, PMT seeing and barometric pressure are all continuously monitored for changes that would affect the data. A sample spectrum is shown in Figure 4, with the Th/Ar fibres above and below the stellar fibre in the centre.

### Deriving the radial velocity

The cross-correlation technique is used to determine radial velocities. As noted, the *IRAF* procedure *FXCOR* was used for this analysis, which also corrects the velocity differences for the orbital and rotational motion of the Earth. Conventionally, digital cross-correlation uses a template star spectrum, such as a star of similar spectral type to the object of study. An alternative technique is to use a synthetic template, such as from a model atmosphere, or even a simple mask using box-shaped spectral lines at appropriate positions.[9] Both methods were employed, *i.e.*: 1) An exposure of tau Boo was used as the template, and all other star exposures were correlated against it; 2) A box-spectrum template, based on a spectrum of tau Boo, was also used to correlate

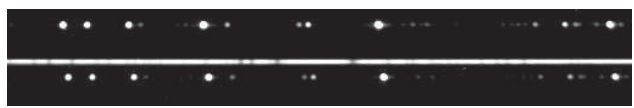

**Figure 4.** Raw CCD image of spectrum. The two thorium/argon reference-lamp spectra are shown in the upper and lower sections while the stellar spectrum is shown in the middle panel.

# Analysis

## Spectral analysis and period determination

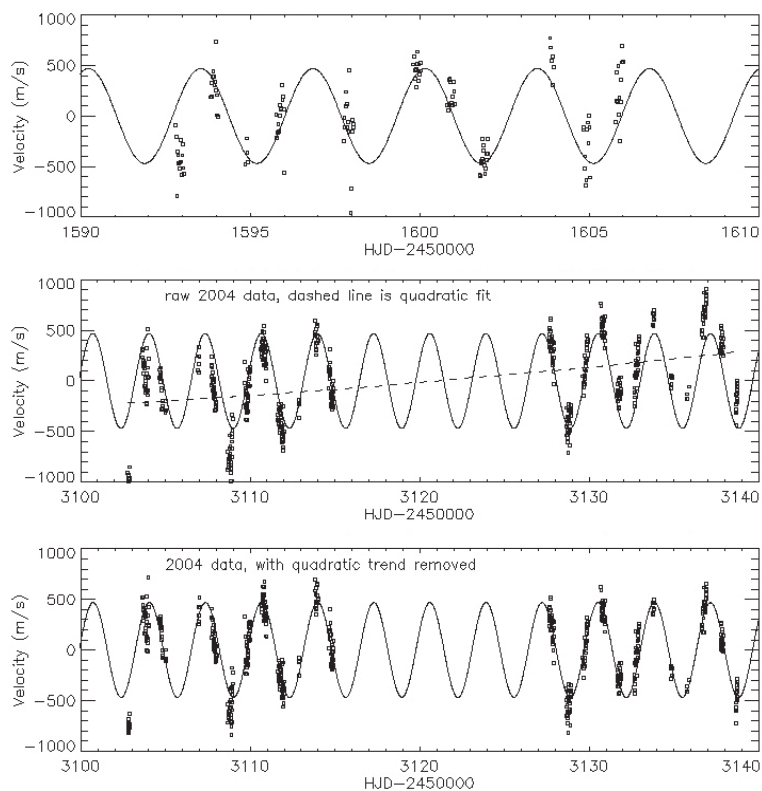

**Figure 5.** Doppler velocities for tau Boötis (points: current data; line: velocity orbit calculated from the results of Butler *et al.*[6]). Top panel: from 2000 Feb 15 to 2000 March 7. Middle panel: 2004 April 6 to 2004 May 12. An upward trend, believed to be due to instrumental effects, is present. A quadratic fit to the trend is shown by the dashed line. Lower panel: 2004 data detrended by the quadratic fit.

against all the star exposures. In large part the results were the same, within observational error, for the two approaches.

Changes in radial velocity were determined with time: the actual, absolute, radial velocity of the star cannot be determined by these methods. Hence the mean of the heliocentric-corrected velocity differences was set to zero for the 2000 Feb–March and the 2004 April–May data sets. As the phase cover was reasonably extensive in each run we believe that this approach is justifiable.

The velocity differences obtained are shown in Figure 5 for the 2000 February–March (top panel) and the 2004 April–May data (middle panel), together with the calculated radial-velocity change of tau Boo based on the Butler *et al.* (1997),[6] Fischer *et al.* (2001),[10] and later Lick results,[11] which yield a period of 3.3125d (solid line in plot). As noted, there are more data from 2004 than for the 2000 run, and the 2004 data generally are of better quality.

The 2000 data, although scattered, show a measure of consistency with the calculated orbit. The 2004 data (middle panel), while showing evidence for a ~3d periodic variation, also exhibit a clear upward trend over time. It is believed that this arises mostly from a small mechanical flexure in the spectrograph that occurred continuously over the course of the 2004 observing runs. The trend was fitted with a quadratic function, shown as the dashed line in the middle panel of the figure, and subtracted to obtain the data shown in the lower panel. The 2004 trend-corrected data show a large measure of agreement with the calculated orbit. A more complete analysis is described below.

A spectral analysis was performed using the Lomb–Scargle normalised periodogram (LNP) method[12] for our combined 2000 and (detrended) 2004 data. The resulting periodogram, in this case found from cross-correlating with a high signal-to-noise-ratio tau Boo spectrum as the template, is shown in Figure 6. The upper panel shows the period spectrum up to 20d. The downward-pointing arrow near 3.3d marks the measured period of 3.3125d determined by earlier workers.[6,10] The lower panel of Figure 6 shows an expanded view of the periodogram near 3d. The downward-pointing arrow is as before. The upward-pointing arrow marks the position of the maximum of a parabolic fit (fitted over the approximate range from 3.2 to 3.4d) at 3.305d. The difference between our result and earlier work is much less than the width of the spectral peak. The fine structure in the peak results from aliasing due to the near 4-year gap in the data set. An attempt to remove the aliasing with the 'CLEAN' technique, where the window function of the data is deconvolved from the data transform,[13] unfortunately met with only limited success. From the work of Schwarzenberg–Czerny,[14] applied to the spectral peak shown in Figure 6, the period error is estimated to be ±0.02d, given the estimated error in a typical data point is just less than 200m s$^{-1}$.

For completeness we also performed a string-length period search using the method of Dworetsky.[15] For the combined 2000+2004 data sets, found from cross-correlating with the synthetic 'box-car' spectrum as the template, again a most likely period was obtained near 3.3d. When we restrict our attention to the observations of the 2004 run, which had the best phase coverage, we obtain the string-length spectrum shown in the upper panel of Figure 7. A phase plot (with arbitrary epoch) using the formal resulting minimum string-length period of P=3.313d is shown in the lower panel. The velocity amplitude of ~500m s$^{-1}$ is consistent with the known amplitude of variation of tau Boo.

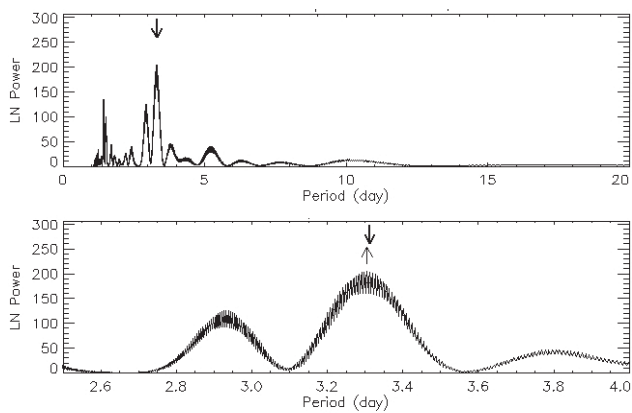

**Figure 6.** Lomb–Scargle normalised periodogram (LNP) of the 2000 and 2004 tau Boo velocity data. Top panel: Periodogram of the data. The downward-pointing arrow indicates the known orbital period of 3.3125d. Lower panel: As above, but showing detail of the spectrum near 3d. The upward-pointing arrow shows the maximum of a parabolic fit to the spectral peak over the approximate range from 3.2 to 3.4d.

The results from our spectral analysis (period = 3.30 ± 0.02d) are consistent with the known period of the extrasolar planet (*i.e.* 3.3125 ± 0.0002d, Butler *et al.*;[6] Fischer *et al.*;[10] Marcy.[11]). The data we have obtained, taken on their own, are unable to refine the period further. However, we can further check for consistency by incorporating the earlier work with our measurements.

Two times of maximum velocity (denoted as $T_{max}$) have been estimated from our data; one each from 2000 and 2004. These were found by fitting a sine wave to each detrended phased data set and determining a time when the sine wave was maximal. The 2000 and 2004 data sets, as HJDs, showed $T_{max}$(2000)=2451600.12±0.1 and $T_{max}$(2004)=2453113.95± 0.01. Taking the $T_{max}$ from Butler *et al.* and the period of 3.3125d,[6, 10, 11] and calculating forward, we can determine *when* velocity maxima would be expected at the epochs of our observations. (This calculation is completely independent of our observations.) Velocity maxima would be expected at HJD 2451600.16 and 2453113.97, in very good agreement with our determinations (viz. 2451600.12 ± 0.1 and 2453113.95 ± 0.01).

If our data are valid, the orbital phase of our $T_{max}$ times must be identical, to within observational error, as a whole number of orbital cycles should have elapsed between them. If we use our $T_{max}$ values (and the errors) with the period of 3.3125d, an estimate can be formed for the number of elapsed cycles as (3113.95–1600.12)/3.3125 = 457.00 ± 0.03, *i.e.* that 457 orbital cycles have elapsed between these data.

It can be concluded that our data are consistent with the findings of the earlier work of Butler *et al.*[6] and Fischer

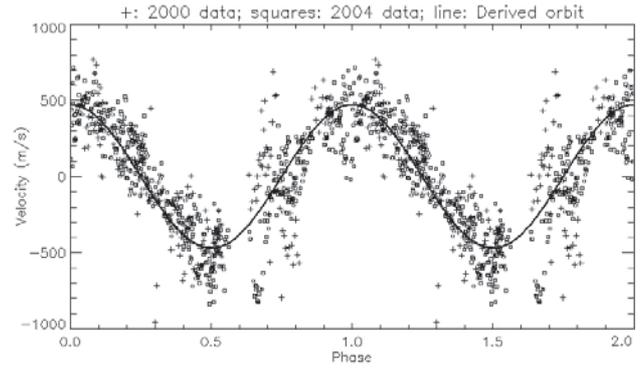

**Figure 8.** Phase plot of the calculated orbit and the combined 2000 and 2004 observations, using a period of 3.3125d. The solid line is the expected velocity orbit calculated from the results derived in this work.

*et al.*[10] We therefore feel confident that our observations have indeed detected orbital motion arising from the planetary companion of tau Boo. Notice that our analysis assumes that only one planet is present in the tau Boo system. Fischer *et al.*[10] have found evidence in their data for small systematic trends in the velocity residuals over an interval of nearly two decades, possibly indicating that at least one additional planet is present.

## Orbital solution

We determined the final orbital parameters of the planet assuming the simple model of a single companion orbiting tau Boötis in a circular orbit. The one-companion model predicts the radial-velocity $V_{RAD}(t)$ of the star at any instant of time. The radial-velocity signal due to the orbital motion of the star with respect to the barycentre of the star and planet system is:

$$V_{RAD}(t) = K\cos[(2\pi/P) \times (t-T_{max})] \quad (1)$$

where $K$ denotes the amplitude of the radial-velocity curve, $P$ is the period of the orbit and $T_{max}$ is the time at which the radial velocity of the star attains a maximum. Equation (1) was fitted to the data by a least-squares procedure while the period was kept fixed at P=3.3125d[6,10,11] (see the previous section) and we obtained for $K$ and $T_{max}$ the optimal values listed in the first column of Table 1. The errors for $K$ and $T_{max}$ were determined by bootstrapping from the residuals of the fit, again holding P fixed at P=3.3125d. This orbital solution is shown as a solid line in Figure 8 with the combined data of the 2000 and 2004 runs. The RMS of the residuals to the orbital fit is just under 200m s$^{-1}$ when a small number of large outliers are removed. It can be observed that the data points follow the derived orbit quite well.

The mass $M_P$ and semi-major axis $a$ of the planet are related to the properties of the orbit by means of the following expressions:[2,5]

$$M_P \sin i/M_J = 0.00493 \times K \times P(days)^{1/3} (M_*/M_\odot)^{2/3} \quad (2)$$

$$a(AU) = 0.0195 P(days)^{2/3} (M_*/M_\odot)^{1/3} \quad (3)$$

where $M_*$, $M_\odot$ and $M_J$ denote the mass of the star, the Sun and Jupiter, respectively, and $i$ is the inclination of the system with respect to the plane of the sky. Note that radial-

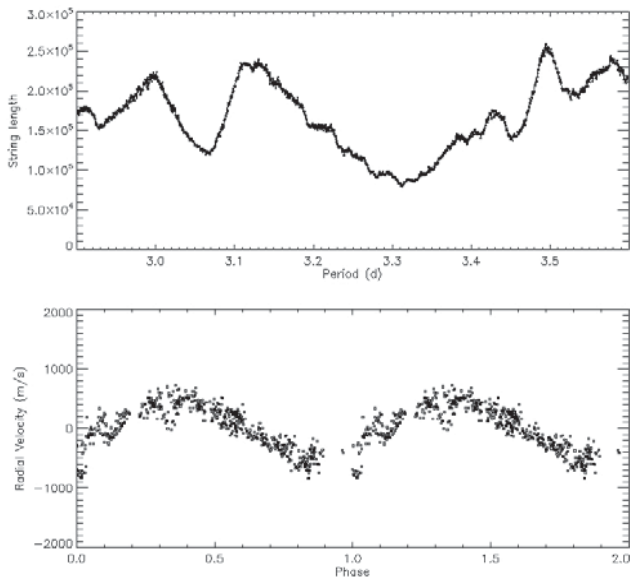

**Figure 7.** Upper panel: String-length spectrum of the 2004 tau Boo velocity data. The minimum string length (best period) is near 3.31d. Lower panel: Phase plot (with arbitrary epoch) of our 2004 tau Boo velocity data using the best period found from the string-length analysis.

Table 1. Comparison of orbital parameters of tau Boötis from the current work and by the 'California and Carnegie Planet Search' team.[6,10,11]

| Current work (eccentricity assumed to be zero) | California and Carnegie Planet Search |
| --- | --- |
| $K = 471 \pm 10$ m s$^{-1}$ | $K = 469 \pm 5$ m s$^{-1}$ |
| $P = 3.3125 \pm 0.0002$ d (adopted) | $P = 3.3125 \pm 0.0002$ d |
| $T_{max} = 2453113.95 \pm 0.014$ | $T_{max} = 2450235.41 \pm 0.2$ d |

Note: The $T_{max}$ values are HJD times of velocity maximum.

velocity measurements can only constrain $M_P \sin i$ rather than $M_P$ alone, and hence provide only a lower limit for the mass of the companion, because the inclination of the system cannot be determined independently with this method. Inserting the relevant values of $P$ and $K$ into equations (2) and (3), and adopting $M_* = 1.2\ M_\odot$ as an estimate for the mass of tau Boötis,[6] we obtain $M_P \sin i = 3.9\ M_J$ and $a = 0.046$ AU.

## Conclusions and future prospects

The radial-velocity measurements presented in this paper clearly demonstrate the feasibility of detecting short-period extrasolar planets with large orbital velocities by the use of telescopes and spectrographs which could, potentially, be built by many dedicated amateur observers and small research institutions around the world.

Spectroscopy in general is an appealing pursuit for astronomers living under light-polluted skies because work can be done 'in between' the light-pollution bands, where the sky is effectively darker. The recent introduction of the commercial SBIG spectrographs is an important step towards implementing spectrographic capabilities on small telescopes. The application of spectral-synthesis techniques to high-quality spectra obtained with those devices would enable quantitative information such as temperature, effective gravity and element abundances to be extracted from stellar spectra recorded with amateur equipment, as is amply demonstrated by the work of amateur astronomer Dale Mais.[16] That field is likely to be a growth area of amateur studies in the years to come.

The detection of the tau Boötis companion, reported here, was initiated as a 'proof of concept', with the goal of discovering an extrasolar planet by non-professional astronomers. Currently the Spectrashift Project (**http://www.spectrashift.com**) is coordinating a team in the design and construction of a 1.1-metre telescope and a high-resolution echelle spectrograph. The telescope will be permanently housed in Arizona and controlled via the Internet. The full-time use of this system will be to search for short-period gas-giant extrasolar planets orbiting nearby main-sequence stars, especially with regard to improving the statistics of extrasolar planets occurring in binaries and orbiting single F-type main-sequence stars.

## Acknowledgments

Tom Kaye thanks Don Barry, Winer Observatory, and the Spectrashift team for their efforts and contributions to the project. Sigfried Vanaverbeke thanks Dr M. Basanze, Prof G. Marcy, Prof C. Waelkens and Prof W. van Rensbergen for their moral support and stimulating discussions on the analysis of radial-velocity curves. John Innis acknowledges his debt to Prof G. Isaak, who passed away while this paper was in preparation, for much patient teaching and for being a continuing source of scientific inspiration. We thank the referee Dr Roger Griffin for his constructive reviews of this paper.

**Addresses: TK:** Spectrashift.com, 404 Hillcrest, Prospect Heights, IL 60090, USA
**SV:** KULAK, E.-Sabbelaan 53, 8500 Kortrijk, Belgium; Vereniging voor Sterrenkunde(VVS), Brieversweg 147, 8310 Brugge, Belgium [Sigfried.Vanaverbeke@kulak.ac.be]
**JI:** 280 Brightwater Rd, Howden, Tasmania 7054, Australia